\documentclass[aps,twocolumn,epsfig,rotate,showpacs]{revtex4}
\usepackage{epsfig}
\usepackage{graphicx}
\usepackage{amsmath}

\newcommand{\beq}{\begin{eqnarray}}
\newcommand{\eeq}{\end{eqnarray}}
\begin{document}
\title{Dissipationless Directed Transport
in Rocked Single-Band Quantum Dynamics}
\author{Jiangbin Gong$^{1}$, Dario Poletti$^{1}$, and
 Peter Hanggi$^{2,1}$}
\affiliation{
$^{1}$Department of Physics and Center for Computational Science and Engineering, National University of Singapore, 117542, Republic of Singapore\\
$^{2}$Theoretische Physik I, Institut f\"{u}r Physik, Universit\"{a}t
Augsburg, D - 86135 Augsburg, Germany}

\begin{abstract}
Using matter waves that are trapped in a deep optical lattice,
dissipationless directed transport is demonstrated to occur if the
single-band quantum dynamics is periodically tilted on one half of
the lattice by a monochromatic field. Most importantly, the directed
transport  can exist for almost all system parameters, even after
averaged over a broad range of single-band initial states. The
directed transport is theoretically explained within ac-scattering
theory. Total reflection phenomena associated with the matter waves
travelling from a tilting-free region to a tilted region are
emphasized. The results are of relevance to ultracold physics and
solid-state physics, and may lead to powerful means of selective,
coherent, and directed transport of cold particles in optical
lattices.
\end{abstract}
\date{\today}
\maketitle

\section{Introduction}
Optical lattices \cite{revmod} have offered new opportunities for
fundamental research in condensed-matter physics
\cite{Jaksch,Paredes,Monteiro06,Eckardt,Wu}. One important example is
Bloch oscillations (BO) \cite{Dahan,Niu,Holthaus} associated with a
periodic potential.  Due to BO, a static bias becomes useless in
generating a net current in the single-band dynamics of a periodic
potential. Hence examining how dissipation helps generate directed
current of  cold atoms/molecules across an optical lattice would
shed light on how electron current gradually emerges from the
interplay of a bias and collision events \cite{kolovsky03a}.

Given this circumstance under which no
directed transport can be coherently generated by
a static bias,
an intriguing question arises:
how can we, if possible, achieve robust directed transport in an ideal periodic
potential
with an oscillating force, in the absence of
any collision effects?
More specifically,  are there simple designs
to realize generic directed
transport
involving only one energy-band ({\it e.g.}, the lowest band)
of a periodic potential, for a broad range of initial states?
Two motivating approaches
attacked this fundamental
question,  but neither of them
was able to reach a very positive and definite answer. In particular,
the first approach directly copes with BO,
with a driving force in resonance with the BO frequency
\cite{zehnle,korsch1,korsch2}. Unfortunately,
the direction of the net transport thus obtained depends
sensitively on the initial state
and on the phase of the driving force.
Hence it is not expected that
the directed transport survives
if the dynamics is averaged over
many initial conditions.
The second approach
relies solely on a driving field that
mixes different harmonics of a fundamental
frequency \cite{goychuk1,goychuk2}.
However, in addition to the requirement
of initial
state coherence (consistent with similar findings in
``coherent control" \cite{brumer}),
the relative phase between
different harmonics should not fluctuate \cite{goychuk2}. If the relative phase
does fluctuate,
then the directed transport was simply transient in the absence of
a bath \cite{goychuk2}, thereby confronted again with
the usage of dissipation to generate current in periodic structures.

Dissipationless directed transport in driven single-band quantum
dynamics, if exists, can be regarded as a type of ``Hamiltonian
ratchet effect" \cite{flach,schanz,gongprl,hanggi,hanggi05}, a
timely topic that attracts great interests recently. Many studies of
Hamiltonian ratchet effects have focused on model systems with
kicking periodic potentials \cite{kick2,kick3,kick3n,kick4,kick5,kick6}.
In these model studies the system is a free particle between
neighboring kicks, hence it is not trapped inside the periodic
potential. As such, if a static bias is allowed to apply to the
system, dissipationless directed transport can easily be generated
in these systems. Conceptually different is the consideration of
Hamiltonian ratchet effect in single-band quantum dynamics, where a
static bias simply does not work. Evidently then, dissipationless
directed transport in driven single-band quantum dynamics, if
established, would constitute a unique class of the Hamiltonian
ratchet effect \cite{flach,schanz,gongprl,hanggi,hanggi05}.

Using matter waves in a deep optical lattice as a possible
realization of a tight-binding model Hamiltonian, we propose in this
paper a straightforward and powerful approach to {\it
dissipationless}, {\it  single-band}, and {\it robust} directed
transport in one-dimensional periodic potentials in the presence of  a
monochromatic driving field. The directed transport results from
fully coherent quantum dynamics associated with a zero-mean driving
field, and is hence unrelated to any sort of system-bath
interaction. Furthermore, the current, irrespective of the details
of system parameters, exists even after averaged over a broad range
of initial states. The results expose a new face of the interplay of
a driving force, energy band properties, and symmetry-breaking in
inducing directed transport. Experimental and theoretical
implications of our finding are vast.

Computational as well as theoretical results also suggest that an
optical lattice with its one half periodically tilted 
carries important applications  for ultracold physics itself. In
particular, total reflection of matter waves travelling from a
tilting-free region to a tilted region is emphasized in this paper.
Such an intriguing aspect of matter waves in an optical lattice can
be very useful for blocking or filtering out one particular
component in a cold gas mixture, an important topic that is attracting
considerable attention \cite{Zoller,Brand}. How
particle-particle interactions might affect the total reflection of
the matter waves in a half-tilted optical lattice will be addressed
elsewhere \cite{gongwork}.

This paper is organized as follows. We first propose in Sec. II our model
system describing matter waves moving in a deep optical lattice, half of which
is subject to a driving field.  This is followed by
computational results that demonstrate the
dramatic consequences due to the driving field.  In Sec. III
we develop a simple scattering theory
to explain and understand the results.  Finally, in Sec. IV we
discuss a subtle symmetry-breaking issue,
compare this work with other related studies of directed transport of
cold atoms, and then draw conclusions.

\section{Matter Waves in a Half-tilted Deep Optical Lattice}

A deep optical lattice can be formed by two interfering and
counter-propagating strong laser beams.  The basic and novel element
here is to periodically tilt one half of an optical lattice.
Although this is experimentally  more demanding than 
periodically tilting the entire lattice via lattice acceleration,
we assume it can be realized and discuss three possible
scenarios.  One possibility is to apply a driving
electric field to one half of the lattice, with the strength of the
electric field linearly changing with the lattice site.
If cold atoms are
in the lattice, then they will experience the static Stark shifts as
a linear function of the lattice site. If cold dipolar molecules are
in the lattice, then the interaction between the electric  dipole
and the driving electric field can give an even stronger tilting
potential. The second possibility is to take advantage of the
magnetic dipole moment of the trapped particles: applying a linearly
increasing magnetic field to one half of the lattice will create a
half-tilted optical lattice as well.  The third scenario is motivated
by the so-called phase imprinting technique in manipulating Bose Einstein
consenates \cite{imprint1,imprint2}.
That is,
an additional far off-resonance 
laser beam covering only one half of
the lattice is applied, with the laser intensity linearly 
varying in space and
periodically modulated.  Such a laser beam interacts with the cold
particles through their induced dipole moment, due to 
the same mechanism as
the optical lattice itself. 

With these considerations,
the quantum dynamics of the cold particle matter wave
can be described by  a tight-binding Hamiltonian
as follows:
\begin{eqnarray}
H&=&-J\sum_{n}(|n\rangle\langle n+1| + |n+1\rangle\langle n|) \nonumber \\
&& +  \cos(\omega t) \sum_{n}nF_{n} |n\rangle \langle n|,
\end{eqnarray}
with
\begin{eqnarray}
F_{n\geq 0}=F, \  \ \ F_{n< 0}=0.
\label{fn}
\end{eqnarray}
Here, $J$ is the tunneling constant (positive) between neighboring
lattice sites, $\omega$ is the tilting frequency of an external
force, and $F$ is the tilting strength of the force.  As clearly
indicated by Eq. (\ref{fn}), only the right half of this lattice is
tilted periodically.  Spatial symmetry is thus broken , but the mean
force is zero. Note also that between the tilted region and the 
tilting-free
region, there is no sudden change in the field stength because
the tilting field linearly increases its strength from zero.
Below we assume $\hbar=1$, and that all systems
parameters are scaled dimensionless variables ({\it e.g.}, the
quasi-momentum of the system will be given in units of $1/d$, where
$d$ is the lattice constant). While focusing on the optical lattice
realization, one should recognize that the above tight-binding
Hamiltonian may be realized in other contexts, {\it e.g.}, electrons
moving in a semi-conductor superlattice with a driving electric
field applied to the right half of the superlattice.

\begin{figure}
\begin{center}
\epsfig{file=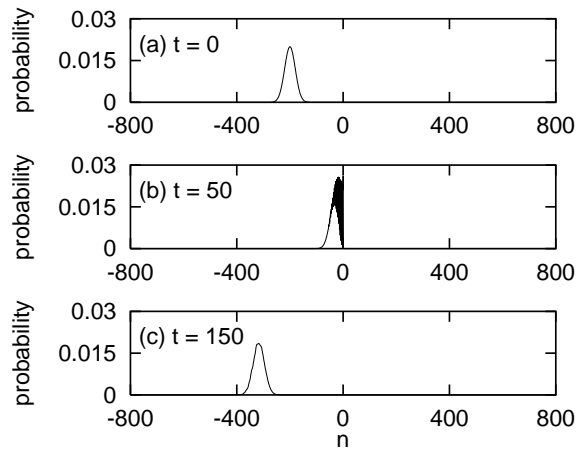,width=8.6cm}
\end{center}
\caption{Complete reflection of a wavepacket travelling from
left (not tilted) to right (tilted).  The system parameters
are $\omega=10$, $F=20$,
$J=2.0$,  and the initial Gaussian wavepacket [see Eq. (\ref{w1})]
has a central
quasi-momentum $k_{1}=\pi/3$, and a position variance $\Delta_{1}=20$.
Panel (a) shows  the initial wavepacket,
panel (b) shows the wavepacket when it is
hitting the $n=0$ boundary of tilting,
and panel (c) indicates that
the wavepacket is bounced back to the tilting-free region.}
\end{figure}

The significant impact of this tilting-half-lattice scenario on the  quantum
transport of cold particles trapped in the lattice   can be first appreciated
by directly examining some wavepacket dynamics calculations.
As one illuminating  example,
consider the case of $\omega=10$, $F=20$, and
$J=2.0$.  The reason why we choose a relatively high driving
frequency $\omega$ is related to a simple scattering theory
developed in the next section (nonetheless, computationally speaking,
using a driving field with relatively low frequencies, {\it e.g.}, $\omega=1.0$,
can generate similar, but less generic results).
The initial wavepacket, denoted $C_{1}(n)$, is given by
\begin{eqnarray}
C_{1}(n)=A\exp(i k_{1} n) \exp\left[-\frac{(n-n_{0})^2}{4\Delta_{1}^2}\right].
\label{w1}
 \end{eqnarray}
Here $\Delta_{1}=20$, $A$ is just a normalization constant.
$k_{1}=+\pi/3$ ($k_{1}=-\pi/3$), and $n_{0}=- 200$ ($n_{0}=200$)
for a wavepacket launched from the left (right) side of
the lattice.

\begin{figure}
\begin{center}
\epsfig{file=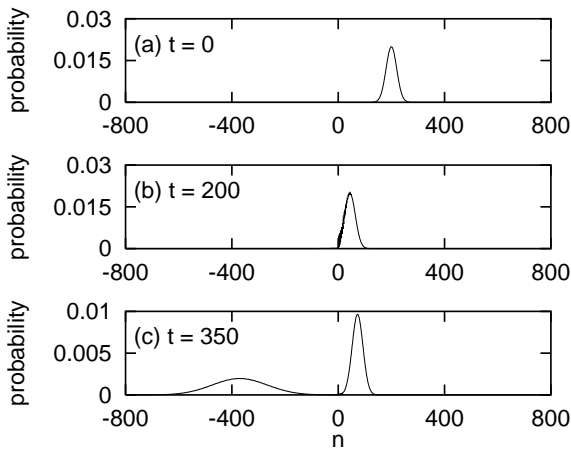,width=8.6cm}
\end{center}
\caption{Significant transmission of a wavepacket travelling from
the right tilted region  to the left tilting-free region. Panels
(a), (b), and (c) show the
wavepacket shape at three different times. Systems parameters are
the same as in Fig. 1, and the initial Gaussion wavepacket is a mirror
image of that shown in Fig. 1a.  At $t=350$,
the probability of finding the atom being at the left half of the lattice
is about 50\%. This should be compared with the total reflection case
seen in Fig. 1.}
\end{figure}

Figure 1 depicts the fate of such a wavepacket initially travelling
from left to right.  At about $t=50$, this wavepacket hits the $n=0$
boundary of the tilting field. Interestingly enough, as manifested
by its location at a later time, {\it e.g.,} at $t=150$, no
wavepacket amplitudes are seen to make their journey all the way to
the right half of the lattice that is being tilted. Instead, the
entire wavepacket is seen to bounce back to the tilting-free region.
The reflection probability numerically calculated is larger than
99.9\%, indicating that this scattering is essentially an event of
total reflection.

In clear contrast, Fig. 2 depicts the  result if an analogous
wavepacket is launched from right to left. The first difference is
that the wavepacket travels at a group velocity much slower than in
Fig. 1. Indeed, only until about $t=200$, does the wavepacket start
to collide with the $n=0$ boundary. But at a later time, about half
of this wavepacket is able to travel across the $n=0$ boundary, and
then continue its travel in the tilting-free region.  The other
amplitudes of this wavepacket are bounced back to the right.

\begin{figure}
\begin{center}
\epsfig{file=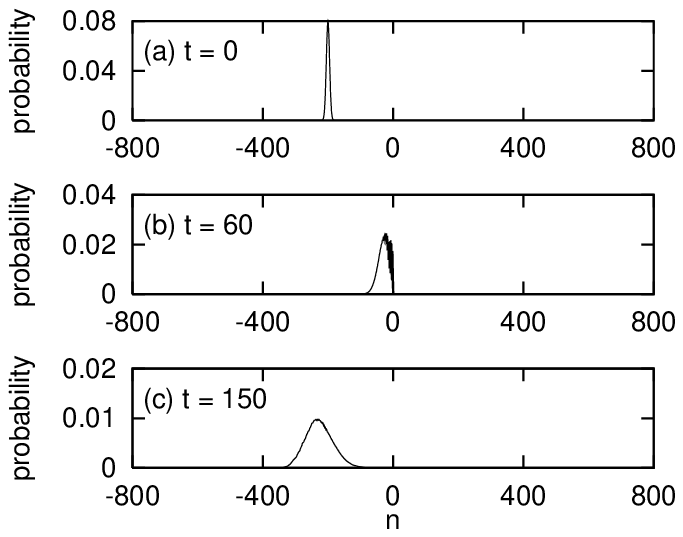,width=8.6cm}
\end{center}
\caption{Complete reflection of a wavepacket travelling from
left (not tilted) to right (tilted).  The system parameters
are $\omega=12$, $F=36$,
$J=2.0$,  and the initial Gaussian wavepacket [see Eq. (\ref{w2})]
has a central
quasi-momtum $k_{2}=0.8$, and a position variance $\Delta_{2}=5$.
Panel (a) shows  the initial wavepacket,
panel (b) shows the wavepacket when it is  hitting the $n=0$ boundary,
and panel (c) indicates that
the wavepacket is bounced back to the tilting-free region.}
\end{figure}

Consider a second sampling case in our wavepacket dynamics
calculations. Here $\omega=12$, $F=36$, and $J=2.0$. The initial
Gaussian wavepacket, denoted by $C_{2}(n)$, is now given by
\begin{eqnarray}
C_{2}(n)=A\exp(i k_{2} n) \exp\left[-\frac{(n-n_{0})^2}{4\Delta_{2}^2}\right].
\label{w2}
\end{eqnarray}
Here $n_{0}$ is the same as before, but we choose $\Delta_{2}=5.0$
to consider much narrower wavepackets as initial conditions.
As for the central quasi-momentum, we choose
$k_{2}=0.8$
for a wavepacket launched from the left side of
the lattice. For a reason to be explained below, which is related to an expression for the group velocity of wavepackets
in the tilted region, we find that
we should still choose $k_{2}=0.8$ (instead of $k_{2}=-0.8$)
to launch an analogous wavepacket travelling from
the right half to the left.

As we deduce from Fig. 3, total reflection of the matter wave also
occurs when the wavepacket travels from left to right. 
Because the wavepacket in Fig. 3(a) has much larger quasi-momentum variance than
that in Fig. 1(a),
its ensuing spreading is also faster. So when this wavepacket hits the boundary [Fig. 3(b)]
it is possible to see
a similar position variance as in Fig. 1(b). 
We
then place this initial wavepacket much closer to the boundary.
Total reflection 
is observed again, and in this case the
position variance at the time of boundary hitting is much
smaller. By contrast,
when an analogous 
wavepacket is launched from right to left (see Fig. 4), significant
probability ($>50$ \%) can eventually be found in the tilting-free
region. These results further confirm that our previous observations made
from  Fig. 1 and Fig. 2  are  general.

\begin{figure}
\begin{center}
\epsfig{file=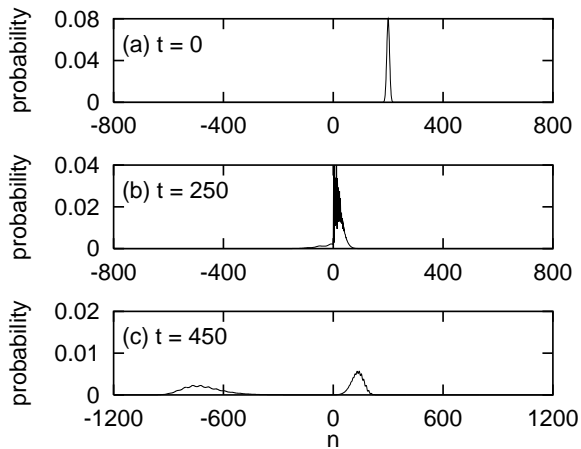,width=8.6cm}
\end{center}
\caption{Significant transmission of an initially narrow
wavepacket travelling from
the right tilted region  to the left tilting-free region. Panels
(a), (b), and (c) show the
wavepacket shape at three different times. System parameters are
the same as in Fig. 3, and the initial Gaussion wavepacket has a
central quasi-momentum $k_{2}=0.8$, and a position variance $\Delta_{2}=5.0$.
The transmission probability is larger than 50\%.
Results here should be compared with those in Fig. 3, where no transmission
is observed.}
\end{figure}

The computational results depicted  and elucidated  above provide a
clear-cut case of symmetry breaking: {\it i.e.}, more particles are
transported from right to left than from left to right. However,
because the details of the wavepacket dynamics depend on the coherence
properties of the initial wavepackets and hence differ from shot to
shot (especially in experiments), the important question is then if
directed transport of cold particles from right to left can survive
when we average the quantum dynamics over a distribution of initial
conditions, and if yes, can we develop a simple theory to identify
the conditions and hence guide the experiments.  This is exactly
what we will elaborate  in the next section.

\section{Simple Scattering Theory}

To rationalize the computational results
we first consider
a well-understood approximation in treating
a globally tilted lattice by an oscillating linear force $fn\cos(\omega t)$
\cite{Monteiro06,Eckardt}. It can be easily shown, even at a
level of classical Hamiltonian dynamics, that
the primary effect of a high-frequency tilting can be accounted for
by re-scaling
the tunneling constant $J$ down to $J {\cal J}_{0}(f/\omega)$,
 where ${\cal J}_{0}$
is the ordinary Bessel function of order zero.
 Formally speaking, this approximation arises from
 a ``$1/\omega$" expansion of an exact Floquet theory of the
 driven quantum dynamics \cite{hanggi}.
In the ``$1/\omega$" expansion of the Floquet theory,
a static Hamiltonian $\tilde{H}$ as the zeroth order approximation
to the Floquet spectrum is given by
\begin{eqnarray}
\tilde{H}=\frac{\omega}{2\pi}\int^{2\pi/\omega}_{0} dt
\exp[i(f/\omega)n\sin(\omega t)]H_{0}\exp[-i(f/\omega)n\sin(\omega t)],
\label{ft}
\end{eqnarray}
where $H_{0}$ denotes the undriven Hamiltonian. In representation
of quasi-momentum $k$,
the tight-binding Hamiltonian of a deep optical lattice can be written as
$H_{0}=-2J\cos(k)$. Then,
using
\begin{eqnarray}
&& \exp[i(f/\omega)n\sin(\omega t)]\cos(k)\exp[-i(f/\omega)n\sin(\omega t)]\nonumber \\
&=&\cos[k+(f/\omega)\sin(\omega t)],
\end{eqnarray}
and
\begin{eqnarray}
\exp[iz\sin(\omega t)]=
\sum^{+\infty}_{l=-\infty}{\cal J}_{l}(z)\exp(il\omega t),
\end{eqnarray}
one immediately obtains that Eq. (\ref{ft}) does yield a
scaling of $J$ by the factor 
${\cal J}_{0}(f/\omega)$.
Clearly,
 this approximation is
 valid if the tilting frequency is high enough. That is, for a large
tilting
 frequency $\omega$,  the probability of finding
the system absorbing (releasing) a net
photon (energy of $\hbar\omega$) from (to) the driving field
in the end
is negligible due to a too large energy exchange.
Then an effective static Hamiltonian $\tilde{H}$
suffices to describe the driven quantum dynamics.
Certainly, within this approach the system is still allowed to
absorb and release an  equal number of virtual photons.

We now adapt this effective Hamiltonian approach to the case of
a half-tilted deep optical lattice.
That is, for the right half of the
lattice, the primary effect of the tilting can be accounted for
by re-scaling
the tunneling constant $J$ down to $J_{R}$, {\it i.e.},
\begin{eqnarray}
J_{R}=J {\cal J}_{0}(F/\omega).
\end{eqnarray}
Note that $J_{R}$
can be negative.
Because the left
half of the lattice is not tilted, the associated tunneling constant,
now denoted
$J_{L}$, is still given by
\begin{eqnarray}
J_{L}= J.
\end{eqnarray}
Given these
considerations, we can describe our system
by effective Hamiltonians $\tilde{H}_{L}$ and $\tilde{H}_{R}$,
for the left and right
halves of the lattice. That is,
\begin{eqnarray}
\tilde{H}_{L} & = & -J_{L}\sum_{n\leq 0}(|n-1\rangle\langle n|
+ |n\rangle\langle n-1|); \\
\tilde{H}_{R} & =& - J_{R}\sum_{n\geq 0}(|n\rangle\langle n+1|
+ |n+1\rangle\langle n|).
\end{eqnarray}
In representation of the associated quasi-momentum $k_{L}$ or $k_{R}$ for particles
moving on the left or on the right,
we have
\begin{eqnarray}
\tilde{H}_{L}&=&-2J_{L}\cos(k_{L}),   \\
\tilde{H}_{R}&=&-2J_{R}\cos(k_{R}).
\end{eqnarray}
Such dispersion relations yield the following group velocities
\begin{eqnarray}
v_{L}&=&2J_{L}\sin(k_{L}),  \\
v_{R}&=&2J_{R}\sin(k_{R}).
\end{eqnarray}
In particular,
the above expression of $v_{R}$ indicates that when $J_{R}$ is negative,
then one needs to have a positive $\sin(k_{R})$ to have a group
velocity in the negative direction. This explains why in the case of Fig. 4
we use
$k_{2}=0.8$, instead of $k_{2}=-0.8$,
to launch a wavepacket from right to left.

The essence of the quantum dynamics for our system
is now reduced to a quantum
scattering problem as a particle travels across two regions with
different dispersion relations.  Great caution, however, is required
because these dispersion relations are distinctively different from
those for free particles.  A trial wavefunction for a left-to-right
scattering event can be written as
\begin{eqnarray}
\psi_{L}(n\leq 0)& = & \exp (ik_{L}n) + r_{LR} \exp(-ik_{L}n); \\
\psi_{R}(n\ge 0) & =& t_{LR} \exp(ik_{R}n) ,
\label{psiR}
\end{eqnarray}
where
\begin{eqnarray}
1+r_{LR}&=& t_{LR},
\label{eq1} \\
J_{L}\cos(k_{L})&=&J_{R}\cos(k_{R}).
\label{energy}
\end{eqnarray}
Considering the sign of the group velocity $v_{L}$, we require
$k_{L}\in [0,\pi]$. Otherwise the group velocity $v_{L}$ of the
incoming wave would be negative, contradicting our assumption.
Analogously,  $k_{R}\in [0,\pi]$ if $J_{R}\geq 0$ and $k_{R}\in
[-\pi,0]$ if $J_{R}\leq 0$. Substituting  $\psi_{R}(n)$ and
$\psi_{L}(n)$ into the discrete Schr\"odinger equation associated with
$\tilde{H}_{L}$ and $\tilde{H}_{R}$, and then
evaluating the coefficient at site $n=0$, we obtain
\begin{eqnarray}
J_{L}\left(r_{LR}-r^{*}_{LR}\right)=J_{R}\left(t_{LR}-
t^{*}_{LR}\right) \exp\left[-i(k_{L}+k_{R})\right].
\label{realeq}
\end{eqnarray}
Equation (\ref{realeq}), together with the condition (\ref{energy}),
suffice to guarantee that $r$ and $s$ are real variables. Moreover,
requiring that the probability at site $n=0$ is  constant, we obtain
\begin{eqnarray}
2J_{L}\sin(k_{L})=2J_{L}\sin(k_{L})r^{2}_{LR}+2J_{R}\sin(k_{R})t^{2}_{LR}.
\label{flux1}
\end{eqnarray}
Indeed, recalling
the group velocities $v_{L}$ and $v_{R}$,
the left hand side of Eq. (\ref{flux1}) is seen to represent
the total
incoming flux, which equals the reflected flux $2J_{L}\sin(k_{L})r^{2}_{LR}$
plus the transmitted flux $2J_{R}\sin(k_{R})t^{2}_{LR}$.

With Eqs. (\ref{eq1}), (\ref{energy}), and (\ref{flux1}), one finds
\begin{eqnarray}
t_{LR}=\frac{2J_{L}\sin(k_{L})}{J_{R}\sin(k_{R})+J_{L}\sin(k_{L})},
\label{trl}
\end{eqnarray}
with
\begin{eqnarray}
k_{R}  =  \arccos\left[\frac{\cos(k_{L})}{{\cal J}_{0}(F/\omega)}\right]
\label{solution0}
\end{eqnarray}
for  ${\cal
J}_{0}(F/\omega)\ge 0 $, and
\begin{eqnarray}
k_{R}  =-\pi+  \arccos\left[\frac{\cos(k_{L})}{{\cal J}_{0}(F/\omega)}\right]
\label{solution}
\end{eqnarray}
for ${\cal
J}_{0}(F/\omega) < 0$.
The same procedure can be applied to right-to-left scattering.
In particular, the analogous transmission amplitude $t_{RL}$ for right-to-left
scattering is found to be
\begin{eqnarray}
t_{RL}=\frac{2J_{R}\sin(k_{R})}{J_{L}\sin(k_{L})+J_{R}\sin(k_{R})},
\end{eqnarray}
with $k_{L}$ given by
\begin{eqnarray}
k_{L}=\arccos[{\cal J}_{0}(F/\omega)\cos(k_{R})].
\label{solutionL}
\end{eqnarray}

With regard to the derivations of the reflection and transmission
amplitudes, additional remarks are necessary. It is very tempting to
apply familiar free-space scattering treatments to the situation
here. For example, one may naively require the derivative
$\partial\psi_{L}(n)/\partial n$ to be continuously  connected with
the derivative $\partial\psi_{R}(n)/\partial n$ at $n=0$. This would
be an incorrect procedure because the connection between the flux
operator and the momentum operator is much different from that in
free space. However, a less rigorous, but enlightening approach in
deriving Eq. (\ref{trl}) does exist by making a more sensible analog
to the familiar scattering theory in free space. Specifically, in
virtue of the fact that the quantum flux operator here is directly
related to $J_{L}\sin(i\partial/\partial n)$ and
$J_{R}\sin(i\partial/\partial n)$,  we have
\begin{eqnarray}
J_{L}\sin\left( i\frac{\partial}{\partial n}\right)\psi_{L}(0)
=J_{R}\sin\left(i\frac{\partial}{\partial n}\right)\psi_{R}(0).
\label{s3}
\end{eqnarray}
With this requirement and Eq. (\ref{eq1})
one can obtain the same scattering results as above.

Intriguing physics can be deduced upon inspecting   Eqs.
(\ref{solution0}) and (\ref{solution}). That is, if the right half
of the lattice is tilted such that
\begin{eqnarray}
\left|\frac{\cos(k_{L})}{{\cal J}_{0}(F/\omega)}\right|>1,
\label{inequality}
\end{eqnarray}
then for such $k_{L}$ there is no solution for $k_{R}$, implicitly
assumed to be real in the trial wavefunction. One might argue that
when a real solution of $k_{R}$ does not exist, then an imaginary
$k_{R}$ could offer a solution describing a state exponentially
decaying in the right half of the lattice. Interestingly, this is
not the case, because the trial wavefunction $\psi_{R}(n\geq 0)$
[see Eq. (\ref{psiR})] with an imaginary $k_{R}$ can never satisfy
the effective, stationary Schr\"{o}dinger equation of the discrete
system here. Clearly, when the inequality (\ref{inequality}) holds,
then $k_{R}$ does not exist and hence $t_{LR}$ must be zero. That
is, no transmission is allowed for the left-to-right scattering,
thereby theoretically confirming our previous observations made from
Fig. 1 and Fig. 3. By contrast, in the case of right-to-left
scattering, for arbitrary $k_{R}$  a solution for $k_{L}$ is
guaranteed [see Eq. (\ref{solutionL})]. This is evident because
$\arccos[{\cal J}_{0}(F/\omega)\cos(k_{R})]$ is always well defined
(note that $|{\cal J}_{0}(F/\omega)|\leq 1$). This identifies a
strongly  broken symmetry, suggesting the possibility of more
particles transported from right to left than transported from left
to right.

To emphasize that the above observation
is a rather general feature for a broad range of
initial states, we now consider the average
transmitted flux $\overline{\Phi}_{LR}$
for left-to-right
scattering. The averaging is over a range $[0, \Delta_{k}]$,
with the convenient assumption
that each quasi-momentum state within this range
has  equal probability.  Then
\begin{eqnarray}
\overline{\Phi}_{LR}(\Delta_{k})
=\frac{1}{\Delta_{k}}\int_{0}^{\Delta_{k}}
t_{LR}^{2}J_{R}\sin(k_{R})dk_{L}.
\end{eqnarray}
In the same fashion,
the average transmitted flux $\overline{\Phi}_{RL}$
for right-to-left scattering can be defined, {\it i.e.},
\begin{eqnarray}
\overline{\Phi}_{RL}(\Delta_{k})=\frac{1}{\Delta_{k}}
\int_{0}^{\Delta_{k}} t_{RL}^{2}J_{L}\sin(k_{L})dk_{R}
\end{eqnarray}
for $J_{R}\geq 0$, and
\begin{eqnarray}
\overline{\Phi}_{RL}(\Delta_{k})=\frac{1}{\Delta_{k}}
\int_{-\pi}^{-\pi+\Delta_{k}}
t_{RL}^{2}J_{L}\sin(k_{L})dk_{R}
\end{eqnarray}
for $J_{R}<0$.

Figure 5 compares $\overline{\Phi}_{LR}(\Delta_{k})$ with
$\overline{\Phi}_{RL}(\Delta_{k})$ as a function of $F/\omega$, for
$\Delta_{k}=\pi/3$, a case representing severe averaging over a
broad range (but still less than half of the entire range) of
initial quasi-momentum states. It is seen that except for
zero-measure cases (also discussed below),
$\overline{\Phi}_{LR}(\Delta_{k})$ is always less than
$\overline{\Phi}_{RL}(\Delta_{k})$. Their difference indicates that,
for  arbitrary tilting frequency $\omega$ and arbitrary tilting
strength $F$, there generically exists a  net transport of particles
from right to left. Even more significant, when $F/\omega$ exceeds a
threshold value ({\it i.e.}, when ${\cal J}_{0}(F/\omega)<0.5$ in
the case of $\Delta_{k}=\pi/3$), then total reflection occurs for
any $k_{L}$ within $[0,\Delta_{k}]$, hence
$\overline{\Phi}_{LR}(\Delta_{k})=0$ whereas
$\overline{\Phi}_{RL}(\Delta_{k})$ can be significant. As seen from
Fig. 5, this leads to a truly dramatic effect with a broad range of
initial states averaged over: only particles launched from the right
can travel to the left, and no particle is allowed to travel from
the left half to the right half.
For these cases, the results in
Fig. 5 are also indicative of how $F/\omega$ must be tuned in order
to generate an optimal transmission flux from right to left.

It should be noted, however, that the zero flux from left to right,
as shown in Fig. 5, is a theoretical result based on a treatment
with the static Hamiltonians $\tilde{H}_{L}$ and $\tilde{H}_{R}$.
The actual flux from left to right might not be mathematically zero,
but should be extremely small. Indeed, in the complete reflection cases
considered in Fig. 1 and Fig. 3 where initial Gaussian wavepackets
are used, the reflection probability never assumes exactly
100\%, but is extremely close to 100\%.

\begin{figure}
\begin{center}
\epsfig{file=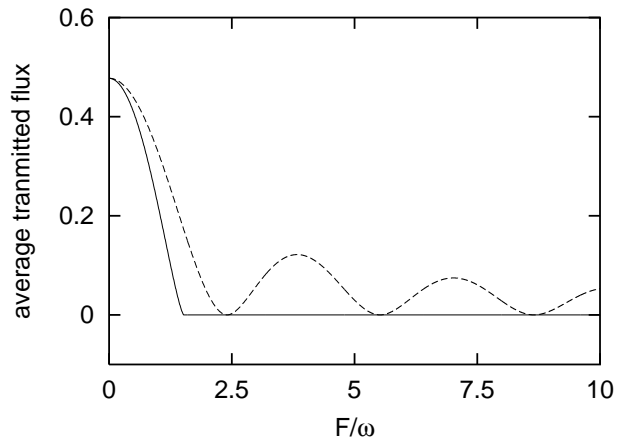, angle=270, width=8.6cm}
\end{center}
\caption{The transmitted flux as a function of
$F/\omega$, averaged over a broad range of
single-band initial states (see the text for details).
Solid line is for left-to-right scattering [$\overline {\Phi}_{LR}(\Delta_{k})$]
and dashed line is for right-to-left scattering [$\overline
{\Phi}_{RL}(\Delta_{k})$], with $\Delta_{k}=\pi/3$.
$\overline {\Phi}_{LR}(\Delta_{k})$ is seen to be generically smaller than
$\overline{\Phi}_{RL}(\Delta_{k})$. Note that
when $F/\omega$ exceeds a threshold value, even
the averaged left-to-right flux is
always zero. The difference between $\overline{\Phi}_{LR}(\Delta_{k})$
and $\overline{\Phi}_{RL}(\Delta_{k})$ gives rise to a net right-to-left
flux of cold particles.}
\end{figure}

Notably, as also elucidated with  Fig. 5,  when ${\cal
J}_{0}(F/\omega) = 0$ for $F/\omega=2.4..., 5.5..., \cdots$, then
both $\overline{\Phi}_{LR}(\Delta_{k})$ and
$\overline{\Phi}_{RL}(\Delta_{k})$ are zero and no directed
transport can  be generated. Indeed, in these cases the
``communication" between the left and right is cut off, as a direct
consequence of tilting-induced localization
\cite{Monteiro06,Eckardt,Drese,hanggi,Grifoni98,CDT1,CDT2}:
particles on the right half cannot even tunnel between neighboring
sites.  A similar situation happens if we apply a static force only
to the right half of the lattice.  Then, particles on the right
cannot travel due to Bloch-oscillations, and particles in the left
half can travel and will be bounced back from the $n=0$ boundary.
Because a nonzero right-to-left transmission is necessary to achieve
directed transport from the right end to the left end of the
lattice, it becomes clear that the directed transported induced by a
half-tilted lattice with ${\cal J}_{0}(F/\omega)\ne 0$ lies in not
only the total reflection in left-to-right scattering, but also in
the significant transmission in right-to-left scattering.

Can we still have  directed transport if we average the dynamics
over all possible single-band initial states? Interestingly, it can
be easily proved that if $\Delta_{k}=\pi/2$ (averaging over a
half-filled band) or $\Delta_{k}=\pi$ (averaging over a completely
filled band), then under the strong assumption that each
quasi-momentum state still has equal probability one obtains
\begin{eqnarray}
\overline{\Phi}_{LR}(\pi/2)&=&\overline{\Phi}_{RL}(\pi/2);
\label{symmetry01}
\end{eqnarray}
and
\begin{eqnarray}
\overline{\Phi}_{LR}(\pi)&=&\overline{\Phi}_{RL}(\pi),
\label{symmetry0}
\end{eqnarray}
both of which result in a {\it vanishing} net flux. Together with
the results shown in Fig. 5, this theoretical result has
implications for experiments. That is,  to observe a net transport
of particles from right to left, one must have a certain degree of
control over how particles are injected into the lattice.  For
example, if particles are injected such that more particles occupy
the states at the bottom of the single-band than other states, then
the result $\overline{\Phi}_{LR}(\pi/2)=\overline{\Phi}_{RL}(\pi/2)$
or $\overline{\Phi}_{LR}(\pi)=\overline{\Phi}_{RL}(\pi)$ becomes
irrelevant. Indeed, for these cases the averaging should be over a
range  $\Delta_{k}<\pi/2$. The associated results are then expected
to be analogous to that seen in Fig. 5 and directed transport of
cold particles can be safely predicted.

The required control of how cold particles should be injected into
the optical lattice suggests that certain degree of spatial
coherence of the initial states is needed in order to observe the
directed transport. As already implied by the results in Fig. 3  and
Fig. 4  where narrow Gaussian wavepackets are considered as initial
conditions, this requirement of initial state coherence properties
can be easily met. Indeed, using an uncertainty relation, one
obtains that as long as the initial wavepacket spans over several
lattice sites, then the variance in the quasi-momentum will be
sufficiently small ({\it e.g.}, $<\pi/3$) to ensure the directed
transport. Fortunately this requirement does not present any
difficulty in today's experiments with cold particles. Indeed,
loading cold atoms into an optical lattice with a particular
quasi-momentum in a particular energy band was achieved
experimentally in Refs. \cite{load1,load2}.

\section{Discussion and Conclusion}
The simple scattering theory in Sec. III explains well our
computational findings. The theory is based upon an  effective,
static Hamiltonian arising from  the zeroth order approximation of a
high frequency $``1/\omega"$ expansion of the exact Floquet theory.
Because the static effective Hamiltonian is always time-reversal
symmetric, one might wonder how it is possible to have directed
transport of cold particles that seemingly contradicts with the
time-reversal symmetry. To clarify this issue, we point out that our
results do not contradict with well-established symmetry
requirements for directed transport. In particular, for a static
Hamiltonian system, one always has \cite{hanggi}
\begin{eqnarray}
\langle n_{L} | U(t)|n_{R} \rangle = \langle n_{R}|U(t)|n_{L} \rangle,
\label{symmetry}
\end{eqnarray}
where $|n_{R}\rangle$ and $|n_{L}\rangle$  are quantum states describing an atom
being localized
exclusively at lattice sites $n_{R}$ and $n_{L}$, and $U(t)$ is the propagator
associated with the static effective Hamiltonian.
Equation (\ref{symmetry}) hence indicates that, due to the time-reversal symmetry, the probability
of transporting a particle exclusively
localized at site $n_{L}$ to site $n_{R}$ is identical
with the probability of transporting a particle exclusively localized
at site $n_{R}$ to site $n_{L}$.   This is exactly
one consequence of Eq. (\ref{symmetry0}).  Specifically,
for these initial states without any
spatial coherence, the initial quasi-momenta fill the entire single-band with
equal probability, therefore a zero net flux is also predicted from
our scattering theory.
This makes it clear that
Eqs. (\ref{symmetry01}) and
(\ref{symmetry0}) originate
ultimately from the time-reversal symmetry of the system.
This leads to the rather formal conclusion that
one prerequisite for
directed transport to occur in our
time-reversal symmetric system
is
a certain degree of spatial
coherence in the initial states.

We now stress again the important advantages afforded by this work
as compared with those in Refs.
\cite{zehnle,korsch1,korsch2,goychuk1,goychuk2}. First, only a
single-frequency driving field is used here, with no special
condition imposed on the driving frequency. Indeed, given the
robustness of our approach, one might conjecture that directed
transport may survive fluctuations in the driving frequency. Second,
because the results depicted in Fig. 5 have been averaged over a
broad range of single-band initial conditions, it becomes clear that
even highly mixed quantum states can generate dissipationless
directed transport. In other words, only very limited 
``quantum
purity" in the initial states is needed to ensure dissipationless
directed transport. These advantages make the tilting-half-lattice
scenario a generic and robust approach for directed transport in
rocked single-band quantum dynamics.

It is also interesting to compare
this work with other related studies
of directed transport of cold atoms in
optical lattices
 \cite{renzoni1,renzoni2,flach2}.  Experiments in Ref. \cite{renzoni1,renzoni2}
used dissipative optical lattices arising from near-resonant laser
beams. To verify dissipationless  current here a far-detuned, and
hence conservative optical lattice is required.  The interesting
recent work of
Ref. \cite{flach2} exploits a harmonic-mixing field, a chaotic
layer, and peculiar features in the Floquet states as system
parameters are suitably tuned, possessing a complex dynamics.
By contrast, in our system
the directed transport, which occurs in wide parameter regimes, is generated
by a regular single-band dynamics.

The ability to induce fully coherent and directed transport  of cold particles
in its lowest energy band
might lead to  building blocks
in constructing atom circuits with unusual characteristics. For
example, the net transport rate here [$\sim
(\overline{\Phi}_{RL}-\overline{\Phi}_{LR})$] is an oscillating function
of $F/\omega$,
instead of being proportional to a ``voltage" $\sim F$.
The revealed simple mechanism
of a quantum ``Maxwell demon" without dissipation also suggests that
cold particles in a mixture
may be selectively transported in a fully coherent
fashion.
Likewise, applying the results to single-band
quantum transport of electrons, new electronic
devices with abnormal current-voltage characteristics and even new designs of
coherent electron pumps become possible.

In conclusion, we show, for the first time, that
dissipationless and generic directed transport can emerge
from single-band quantum dynamics driven by a monochromatic
field, even after averaged over a broad range of initial states.
The underlying mechanism of the directed transport is related to
total
reflection vs significant transmission
as the matter wave in a half-tilted optical lattice moves
in opposite directions.
The results
are of fundamental interest to
solid-state physics and ultracold physics.
Experiments using
cold atoms/molecules in a deep  and half-tilted
optical lattice should be able to verify the results of
this study.

{\bf Acknowledgments}: The authors thank the referee for
suggesting the usage of a laser beam with varying laser intensity
to realize a half tilted optical lattice.
J.G. is
supported by the start-up funding (WBS No. R-144-050-193-101 and R-144-050-193-133), National University of Singapore,and 
the ``YIA" funding (WBS No. R-144-000-195-123) from the office of Deputy President (Research \& Technology), National University of Singapore.
J.G. thanks Dr. Jiao Wang, Dr. Wenge Wang and
Prof. Baowen Li for interesting and useful discussions. 
P.H. acknowledges support by the DFG, via the
collaborative research grant SFB-486, project A-10 and the
Volkswagen foundation.

\end{document}